# The digital footprint of innovators: Using email to detect the most creative people in your organization

Gloor, P. A., Fronzetti Colladon, A., & Grippa, F.

**The digital footprint of innovators: using E-Mail to detect the most creative people in your organization**

Gloor, P. A., Fronzetti Colladon, A., & Grippa, F.

**Abstract**

We propose a novel method for finding the most innovative people in an organization, using E-Mail to analyze structure and dynamics of the organization's online communication. To illustrate our approach, we analyzed the e-mail archive of 2000 members of the R&D department of a US multinational company. We use metrics of social network analysis extended with meta-data of interaction dynamics to calculate features for individual employees: their network positions, messages sent and received, pings to others and response times. We find a distinction between innovation group leaders and subject matter experts focused on publishing papers and patents. Innovation administrators have a higher number of direct contacts, are more committed in conversations and receive more messages than they send. We also found significant differences between innovators oriented towards internal awards and innovators more concerned with external recognition of their work.

# 1. Introduction

Measuring creativity of employees has been an active area of research for a long time (Amabile, 1996; Hughes, Lee, Tian, Newman, & Legood, 2018; McKay & Kaufman, 2019). Creativity has been traditionally measured through the administration of individual questionnaires, self-ratings or task assessments. Recent studies have demonstrated how conceptualizing creativity and innovation using different approaches leads to completely different ways of measuring them (Rauner, 2019; Sternberg, 2018). Most of the traditional creativity measures follow a unidimensional model, based on the assumption that creativity can be assessed on a single interval scale, much as intelligence (Frey, 2018; Sternberg, 2006; Torrance, 1972). Several methods to measure creativity have been used in the past, and could be divided into questionnaire-based and task-based methods. These approaches include: personality inventories; questionnaire-based measures; self-reported creative activities; thinking styles inventories; task-based measures; divergent thinking tasks; artistic and real-life creativity tasks and insight tasks (Fürst, 2018; Fürst & Grin, 2018). More research is still needed to explore the interdisciplinary nature of creativity and its impact on individual and organizational outcomes (Brem, Puente-Diaz, & Agogué, 2016). Exploring the connections between communication behaviors and innovation within organizations is not a new topic for researchers (Allen, Gloor, Fronzetti Colladon, Woerner, & Raz, 2016; Ebadi & Utterback, 1984; Fosfuri & Tribó, 2008; Raz & Gloor, 2007; Sousa & Rocha, 2019). What has been less frequently studied is the identification of specific characteristics in online communication to differentiate innovators from non-innovators. What can we learn from the online communication behaviors of engineers and researchers in terms of their innovative potential? Do they interact differently with colleagues and share information at a different pace? To address these questions, we developed a case study with an embedded population of about 2,000 employees working within the R&D department of a global energy company. We analyzed their email communications, i.e. more than 2 million emails in the second quarter of 2016. This study investigates communication behavior of individuals charged with innovative tasks using online social network analysis. In this setting, emails represent a primary communication channel. Despite the availability of competing technologies, emails remain a crucial source of enterprise information and serve 'as a virtual extension of the users' workplace' (Scerri, 2013). In dynamic research environments, the email collaboration space can often be more stable than physical locations of workers, which might frequently change due to research missions or dynamically assigned positions in shared workspaces.

To study the online communication behavior of R&D employees, we relied on existing literature exploring the distinctive individual characteristics and information processing habits of innovative individuals (Amabile, 1988; Dyer, Gregersen, & Christensen, 2011; Keller & Holland, 1983). By mapping email-based social networks and extracting metrics of network structure and interaction dynamics, we investigate what distinguishes different types of innovators involved in the process of new product development. We then explore how their online communication behaviors differ from the behaviors exhibited by other engineers and R&D professionals and from innovation administrators, whose role is to offer suggestions, ideas and support to their team.

The remainder of the paper is structured as follows. The next section provides an overview of the literature on traits and communication patterns associated with innovators, discussing how creativity relates to personality, education/training, and time management. Section 3 describes the research model and defines the hypotheses along with the adopted metrics. Section 4 presents the main results highlighting the differences in communication styles between innovation administrators, innovators recognized for their scholarly publications/patents, and innovators whose ideas were awarded with an institutional prize. Sections 5 and 6 discuss the main implications of the results and conclude with some caveats to our analysis, and a discussion of how our findings can inform managerial decisions in terms of human resource management and organizational design.

## 2.   Traits and Behaviors of Creative Individuals and Innovators

Researchers have been interested in studying behaviors of creative and innovative individuals for at least four decades (Allen, 1977; Amabile, 1988; Dyer et al., 2011; Fürst & Grin, 2018; Kandemir & Kaufman, 2019; Keller & Holland, 1978; Marrone, 2010). As suggested by Sternberg (1985, 2018), different kinds of creativity are associated with different intellectual skills depending on the ability to see a problem in new ways and to depart from conventional thinking. Other individual factors impacting creativity include the skill to recognize which of one's ideas are worth pursuing, and how to convince others of the importance and value of one's ideas (Snyder, Hammond, Grohman, & Katz-Buonincontro, 2019; Sternberg, 2006). Recently, Woo, Keith, Su, Saef and Parringon (2017) and Fürst and Grin (2018) showed that interests in aesthetics, imagination and reflection, as well as openness and intellect are correlated with creative activities, potential and outcomes. Additional personal factors which influence creativity include cultural differences, as demonstrated by different beliefs among Chinese and US individuals (Niu & Kaufman, 2013). For example, humor has been described as a component of creativity in Western cultures, whereas it is less likely to be perceived as such in Eastern cultures (Sternberg, 2003).

As suggested by recent studies measuring intrapreneurship at the individual level (Gawke, Gorgievski, & Bakker, 2019), employee innovative behavior is characterized by a strategic mindset, a strong awareness of external trends and venture activities aimed at creating, adding, or investing resources in new businesses (Gawke et al., 2019; S. H. Park, Kim, & Krishna, 2014). Other specific traits such as personal initiative and social competence have been identified as typical qualities of creative individuals and innovators. For example, Keller and Holland (Keller, 2017; Keller & Holland, 1978) found that communicators/innovators are characterized by similar traits and behaviors, including an innovative orientation, high self-esteem, more formal education, a high degree of reading, supervisory duties, and centrality in communication networks. The authors also found that physical propinquity was associated with the creation of strong connections among communicators, innovators, and other employees, which is aligned with the work of Allen (1977) on physical proximity and innovative outcome. In a recent study involving 274 knowledge workers in 27 small and medium-sized enterprises, Dul, Ceylan and Jaspers (2011) found that creative performance was affected by creative personality as well as by proximity in the physical work environment.

Kaufman and Beghetto (2009) have suggested a Four C model of creativity to explain differences in outcomes of creative endeavors. Their model expands on the concept of everyday creativity (often called "little-c") and goes beyond the so-called eminent creativity, or "Big-C", which is associated with a more significant outcome (e.g. winning a Pulitzer prize). The Four C model is based on the assumption that nearly all aspects of creativity can be experienced by nearly everyone (Kaufman & Beghetto, 2009). Recent studies have incorporated this model of individual creativity and tested its generalizability in different cultural contexts, specifically among people from Germany and Mexico (S. Kim, Choe, & Kaufman, 2019; Puente-Diaz,

Maier, Brem, & Cavazos-Arroyo, 2016). A recent study by Rinne, Steel and Faitweather (2013) explored how the creative process, its output, as well as the societal perception of this output, are all influenced by the cultural context in which the creative process takes place.

In another recent study, Dyer et al. (2011) found that innovative entrepreneurs spend 50% more time on discovery activities than do CEOs with no track record for innovation. In three different studies involving R&D scientists, Amabile (1988) examined different factors influencing creativity and innovation in organizations. In addition to specific personality traits associated with innovative behaviors (e.g. persistence, curiosity, energy, and intellectual honesty), the study also found that possessing good social and/or political skills, good rapport with others, and being a good listener and a good team player, were all very important behaviors for an innovator (Amabile, 1988). In their study, Ray and colleagues (1997) found that organizational innovators were more excited to communicate with others, and were less apprehensive about a variety of communication situations compared to their colleagues.

Recently, Sauerman and Cohen (2010) found that innovation – measured by the number of individuals' U.S. patent applications – was positively associated with intellectual challenge, independence, and monetary reward.

Daft (1978, 2007) suggested a theory based on a dual-core model, differentiating between technical innovation, namely a bottom-up process initiated and executed by lower-level employees with technical expertise, and administrative innovation, which falls in the area of administrators or upper-level managers. Technical innovations include the introduction of new products, services, and production process technologies, while administrative innovations are more directly related to its internal management (Damanpour, Sanchez-Henriquez, & Chiu, 2018). This distinction is aligned with our differentiation between product-oriented innovation, award-oriented innovation, and innovation administrators. While in the first two categories we find researchers focused on technical innovation, in the third one we find upper-level managers who are aligned with administrative innovation. Our two categories are based on the assumption that different types of innovations are associated with different demands and constraints, require different decision processes, and thus may require the mobilization of different sources of power (Ibarra,1993). Innovation administrators are the facilitators of the creative process, the mentors who support with their expertise and knowledge both award and product oriented innovators. They are involved in administrative tasks to promote innovation, although they are not directly asked to write scientific papers, submit patent applications or design prototypes for the institutional award. Innovation administrators are the subject matter experts, working as project coordinators with supervisory roles, and acting as a point of reference for other researchers. In the past they might have been product-oriented or award-oriented innovators themselves. Once their career advanced, they became innovation administrators, switching to a role that is less operative and more administrative. This distinction is based on traditional measures of R&D effectiveness which are built on quantitative indicators such as the number of published papers, number of patents filed, successful technology transfers and grants associated to the R&D team. More recently, composite quantitative metrics have been used, such as the impact factor of the journals where papers appear and the science citation index (B. Kim & Oh, 2002; Roy, Nagpaul, & Mohapatra, 2003).

## 2.1. Hypotheses

Given the importance of collaborative relationships to act as critical conduits for knowledge search, transfer, creation, and innovation, R&D organizations and teams rely on individuals who can help build a network of linkages and offer access to valuable sources of knowledge both within and outside the organizational boundaries (Cross & Prusak, 2002). Literature commonly refers to this phenomenon as 'team boundary spanning' (Marrone, 2010): individuals who span organizational boundaries act as important translators for external knowledge as they make it available and relevant to the unique requirements of an organization (Fleming & Waguespack, 2007). Successful innovations seem to be dependent on the ability to cross boundaries, accessing new ideas and nurturing previous collaborations (Cowan, Jonard, & Zimmermann, 2007).

Innovators question, observe, experiment, and network more than typical executives (Dyer et al., 2011). As described by Dyer et al. (2011), the innovator's DNA is based on five discovery skills: associating, questioning, observing, experimenting, and networking. Innovative entrepreneurs constantly ask questions that challenge assumptions and conventional wisdom. They invest time and energy to developing and nurturing new connections outside of their environment, finding ideas through a network of diverse individuals that will expose them to different perspectives. Innovators are often reported as possessing good interpersonal skills, as being able to develop numerous contacts with others and act as boundary spanners (Amabile, 1988; Dyer et al., 2011; Perry-Smith, 2006; Ray et al., 1997).

Recent studies indicate that centralization may be necessary for the effective implementation of ideas and the commercialization phase (West & Richter, 2008), though it could be detrimental in the idea generation stage. As Becker noted (1970), centrality in a social network of individuals who communicate and exchange knowledge can be associated with innovation capabilities: innovators can benefit from their network position by gaining access to several sources of information, which can provide legitimization and support for an innovation. Given the strong connection between individual network centrality and source of power (Brass, 1984; Tsai, 2001), we would expect innovators' online communication to be characterized by high degree centrality.

*H1: Innovators have more direct contacts in their network, i.e. a higher degree centrality.*

Other studies explored the role of social networks in fostering creative outcomes, looking at core-periphery network structures, where core actors are very central while those in the periphery are less central, to explain award allocation decisions. In a study of the social connections within the Hollywood movie industry, Cattani and Ferriani (2008) found that individuals who occupy an intermediate position "between" the core and the periphery of their social system are likely to achieve creative results: "*By being close to the core, they can benefit from being directly exposed to sources of social legitimacy and support crucial to sustaining creative performance; at the same time, by not losing touch with the periphery, they can access fresh new inputs that are more likely to blossom on the fringe of the network while escaping the conformity pressures that are typical of a more socially entrenched field*" (Cattani & Ferriani, 2008, p. 838). Other scholars focused on the social determinants of creativity (Amabile, 1996; Csikszentmihalyi & Sawyer, 1995) and emphasized how innovators need to be recognized by a community that endorses the quality of their contribution in order to be successful in their creative endeavor. As Burt (2004) suggested, a successful ideation process depends on the ability to broker knowledge and bridge structural holes across various social networks. This leads to the following hypothesis:

*H2: Innovators act as information hub in their network, having a higher betweenness centrality.*

The results of a qualitative study involving 32 organizational members  provided evidence that innovators tend to be more eager to communicate, are less apprehensive about communicative events and show a desire to talk for longer periods of time (Ray et al., 1997). Other studies indicate that innovators are perceived as good listeners and good team players (Amabile, 1988; Gawke et al., 2019). Innovators thrive in an organizational culture that supports autonomy, experimentation and engagement (Chesbrough, 2006; Gagné & Deci, 2005). Successful innovators are more passionate about conversation and more curious to learn from others, and this enthusiasm should be visible in their interactivity level. It is not enough to have big ideas: successful innovations come from clear and frequent communications with different stakeholders, to clarify expectation, refine details and push the idea to the next level. Instead of sitting alone at their desks, innovators spend a lot of time and effort communicating and collaborating with others (Paulus & Nijstad, 2003). While careful listening and team playing are maximized during face-to-face meetings, we maintain that such an attitude of innovators will be at least partially reflected in their email communication. As suggested by recent studies on virtual research teams (Lee & Bozeman, 2005; Walsh & Maloney, 2007), access to emails reduces coordination problems and increases productivity, though some differences exists between developed and developing country scientists. In our research setting, not every team was co-located and email was often a preferred mean of communication (when not using online conference services). Past research showed that researchers who were located in developed countries, had immediate access to email technology without any local variation in connectivity and were expected to have a quick turnaround (Duque et al., 2005; Sooryamoorthy, Duque, Ynalvez, & Shrum, 2007). There was no difference in corporate culture regarding email practices. In their qualitative study on email responsiveness involving two large global tech companies, Tyler and Tang (2003) found that participants expected a quick response (under an hour) based on how quickly they had responded in the past, and that they formed this expectation after just a few interactions. More importantly, they found that respondents displayed typical patterns of response behaviors, including maintaining a responsiveness image and reciprocate the email behavior of others: the more responsive people are, the higher the chance receivers will conform to this behavior. Of course, the expectation of responsiveness varies from person to person, but in a team where individuals work together toward the same common goal, we would expect individuals who feel the same commitment and sense of urgency, will likely be more responsive. More recently, Tinguely, Ben-Menahem, He and von Krogh (2019) found that employees with a higher position within the organization and high relational resources experience less energy depletion when time pressure rises over time. Based on this reasoning and past empirical evidence, we would expect that innovators are likely to respond to other people's emails more promptly (H3), and to be more engaged in conversations, frequently pinging others to stimulate the discourse (H4). Therefore, we would expect that:

*H3: Innovators tend to respond to emails faster.*

*H4: Innovators show a higher engagement in the conversation, pinging others more frequently.*

As suggested by studies on the contribution of individuals in online communication networks (Gloor, 2005, 2016), creators and innovators tend to have a balanced exchange of received and sent messages. Actors who only send messages, without receiving any, are spammers in most of the cases (Fronzetti Colladon & Gloor, 2019), whereas there are also actors who are often mentioned and copied in emails who do not often answer back; these people might be experiencing a disengagement with their job (Gloor, Fronzetti Colladon, Grippa, & Giacomelli, 2017); or by contrast they could be very senior employees, with supervisory duties, thus without the need to answer frequently to day-by-day activities. Based on several studies about the balance in communication for an effective knowledge sharing (Gloor, 2005, 2017; Gloor, Paasivaara, Schoder, & Willems, 2008), innovators seem to be positioned in the middle, with similar values of messages sent and received. The literature mentioned above supports our fifth hypothesis:

*H5: Innovators maintain a balance between messages sent and messages received.*

The processes required to ensure the success of an entrepreneurial activity within an organization tend to involve similar practices, such as leveraging social capital and political intelligence. What could change is the specific information or resource that innovators are trying to find in the networks. For example, award-oriented innovators will try to connect with others within the organization to find potential supporters of their idea. Product-oriented innovators could seek advice on how to better describe the innovative idea of their patent application. Whatever information or support innovators are looking for, they will rely on their social networks and take advantage of their position regardless of the type of innovation they are promoting (Ibarra et al., 1993; Kanter, 1988). We do not expect to see a significant difference in communication behavior between product-oriented innovators and award-oriented innovators in terms of centrality or interactivity, since the outcome of the innovation process should not impact the inclination of innovators to connect with others to promote the diffusion of their ideas via patents, publications or prizes. Several empirical evidences suggest that informal networks play an important role in promoting any type of innovation, including administrative and technical innovation, process and product innovation (Frost & Egri, 1991; Ibarra et al., 1993; Lumpkin & Dess, 2001; Tsai, 2001). These studies seem to suggest that the only difference in communication behavior is between employees holding an administrative role and employees whose job is to invent the next "big thing". Therefore, we propose the sixth and final hypothesis:

*H6: There is not a significant difference in online communication behavior of product-oriented innovators and award-oriented innovators.*

## 3. Research Method and Variables

We analyzed the email communication within a global energy company. During the second quarter of 2016, we collected a sample of more than 4 million emails (due to privacy agreements, without bodies and subject lines), corresponding to the inboxes of about 4,800 people working in the R&D department. From this original sample, we removed the employees who were operating in locations outside the US, to create consistency in terms of culture, internal rules and local team size; we also removed external email accounts, such as the ones belonging to trainees or suppliers. The final sample was composed of 1944 internal employees who exchanged more than 2 million emails. When extracting social network metrics we also recorded the communication exchanges that crossed the company's boundaries. In the network graphs, nodes were representing email accounts, with an arc originating at node A and terminating at node B, if A sent B an email.

Out of the 1944 employees included in our sample, 211 researchers have been identified as highly productive innovators. Among these 211 researchers involved in innovative projects, 54 had previously been recognized by leadership with a prestigious internal innovation award, while 131 were identified as product-innovators, i.e. employees who had filed patents or published scholarly articles, and 26 had been designated by leadership as innovation administrators. As mentioned in the introduction, innovation administrators are the subject matter experts, employees who advanced their career thanks to their knowledge and management skills; they now have supervisory roles, and coordinate the work of the other researchers, coaching them and sharing advice. An internal committee gave the awards to those employees who proved their commitment to the research objectives of the company and who showed outstanding innovation skills, providing a significant contribution to the research projects they participated in (they might for instance be working on projects where publishing, or applying for patents, was not possible). Different metrics of innovativeness reflect different kinds of creative performance: whereas patents or academic publications represent a radical type of creative performance, awards reflect an adaptive kind of creativity outcome, as the judges would review and reward results that were considered original and suitable for organizational implementation (Oldham & Cummings, 1996).

We used metrics of Social Network Analysis (Wasserman & Faust, 1994) to evaluate employees' degree of *connectivity* – which measures how central and well connected individuals are – and degree of *interactivity* – which measures the tendency of individuals to send and respond in a timely manner to other people's emails (Gloor, 2005). From an interpersonal communication perspective, interactivity is defined as the ability of two individuals to communicate directly with one another, regardless of distance or time. To measure interactivity, we calculated the following metrics: average response time (ART); nudges – consisting in the number of follow ups necessary to prompt a response from another person, as well as to the number of pings to others in a conversation with multiple actors; number of messages sent; number of messages received; the balance in messages received and sent (Gloor, 2016). To measure connectivity we used two well-known social network metrics, betweenness centrality and degree centrality (Wasserman & Faust, 1994). Degree centrality represents the number of direct contacts an employee has. Betweenness centrality, on the other hand, counts how many times an actor lies in-between the shortest network paths that interconnect his/her peers. This metric is often used as a proxy of brokerage power of social actors, i.e. their ability to influence or control information that goes beyond direct links (Wasserman & Faust, 1994). All the metrics used in this study are described in Table 1. For all the variables we considered the percentile rank scores, instead of their absolute values, to better determine high and low values in relation to their frequency distributions.

| Metrics | Definition | Operationalized as |
|---|---|---|
| **Degree Centrality** | Number of actors each person is directly connected within a network | Number of direct contacts of an actor, both as senders or receivers in the network |
| **Betweenness Centrality** | The extent to which each actor acts as an information hub and controls the information flow | It is defined as the likelihood to be on the shortest path between any two actors in the network |
| **Ego Average Response Time (ART)** | Average number of hours sender takes to respond to e-mails | Time until a frame is closed for the sender, after he has sent an e-mail |
| **Alter Average Response Time (ART)** | Average number of hours receiver takes to respond to e-mails | Time until a frame is closed for the receiver, after he has received an e-mail |
| **Ego Nudges** | Average number of follow-ups that the sender needs to send in order to receive a response from the receiver | Number of pings until receiver responds |
| **Alter Nudges** | Average number of follow ups that the receiver needs to send in order to receive a response from the sender | Number of pings until sender responds |
| **Messages Received minus Sent** | Indicates how balanced a communication is in terms of messages received and sent | Messages Received-Messages Sent |

**Table 1.** Metrics used to measure degree of connectivity and interactivity.

Lastly, we considered employees' *rank* within the organization as a control variable and included it in our models (a score of 1 corresponds to the highest rank and a score of 3 to the lowest one). Indeed, we expect employees with a higher rank to be more central, as their hierarchical position often implies having more people to supervise. This variable can also have a significant association with the fact of being an innovation administrator, as these people are often promoted to higher ranks.

Figure 1 summarizes the hypotheses and connects them to the research model.

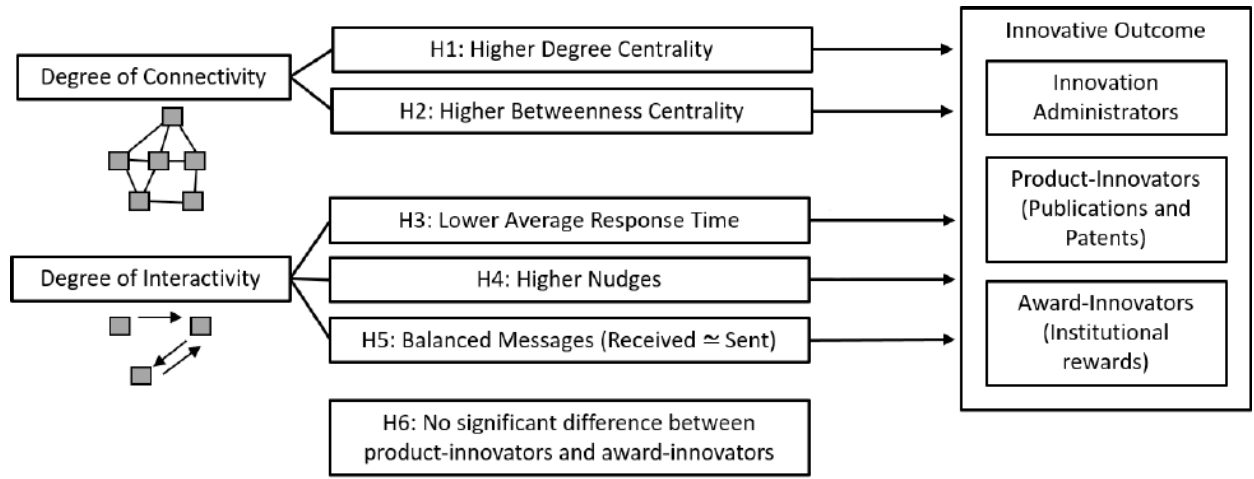

**Figure 1.** Hypotheses.

## 4. Results and Discussion

The first step in our analysis was to explore how innovation administrators communicate via email. As the logit models of Table 2 show, the most important predictor to identify innovation administrators is ego nudges, representing the average number of follow-ups that the sender needs to send in order to receive a response from the receiver. It seems that senior experts pinged more frequently, which could be associated to a greater commitment to the discussion and a higher level of engagement as found in other studies on measuring response time and engagement level of employees (Gloor, Fronzetti Colladon, Grippa, et al., 2017; Tyler & Tang, 2003). At the same time, innovation administrators are characterized by higher degree centrality and an unbalanced pattern of messages received vs sent. Innovation administrators receive more emails than they send, which seems consistent with their role as a point of reference for others whenever their expertise is needed. They also tend to hold a central and prominent role in their email network as indicated by their high degree centrality (Everett & Borgatti, 2005; Wasserman & Faust, 1994). We additionally controlled for rank, to exclude the effects that higher hierarchical positions can have on centrality metrics and on the probability of being an innovation leader. The significance of our predictors and a value of the McFadden's $R^2$ of 0.224 in the final model, indicate a good fit.

| Variables | Model 1 | Model 2 | Model 3 | Model 4 | Model 5 | Model 6 | Model 7 |
|---|---|---|---|---|---|---|---|
| Rank | -1.375*** | | | | | -0.719* | -0.729* |
| Ego ART | | 1.194 | | | | | |
| Alter ART | | -0.546 | | | | | |
| Ego Nudges | | | 2.612** | | | 3.717** | 3.745*** |
| Alter Nudges | | | 3.045*** | | | 0.178 | |
| Messages Received-Sent | | | | 3.757*** | | 3.606*** | 3.683*** |
| Degree Centrality | | | | | 4.888** | 4.821** | 4.798** |
| Betweenness Centrality | | | | | 0.1304 | | |
| Constant | 2.678568 | -4.667 | -7.598*** | -6.693*** | -8.168*** | -8.921** | -8.817** |
| McFadden's $R^2$ | 0.078 | 0.009 | 0.073 | 0.076 | 0.059 | 0.224 | 0.224 |
| N | 1944 | 1944 | 1944 | 1944 | 1944 | 1944 | 1944 |
| AIC | 258.488 | 279.469 | 261.808 | 259.131 | 265.734 | 226.214 | 224.235 |
| BIC | 269.633 | 296.186 | 278.526 | 270.276 | 282.451 | 259.649 | 252.098 |

***p<.001;**p<.01; *p<.05

**Table 2.** Communication style of innovation administrators.

Similar results were obtained in the t-tests of Figure 2. We notice some significant differences in the average behavior of innovation administrators when compared to their peers. All the metrics seem to indicate a specific communication pattern for senior experts, with the exception of average response time (ART). Innovation administrators are confirmed as more central in their email communications than the rest of their colleagues, particularly in terms of the number of other actors they are connected to. They also receive more messages than they send and participate in more engaged conversations, showing higher values of Ego and Alter Nudges.

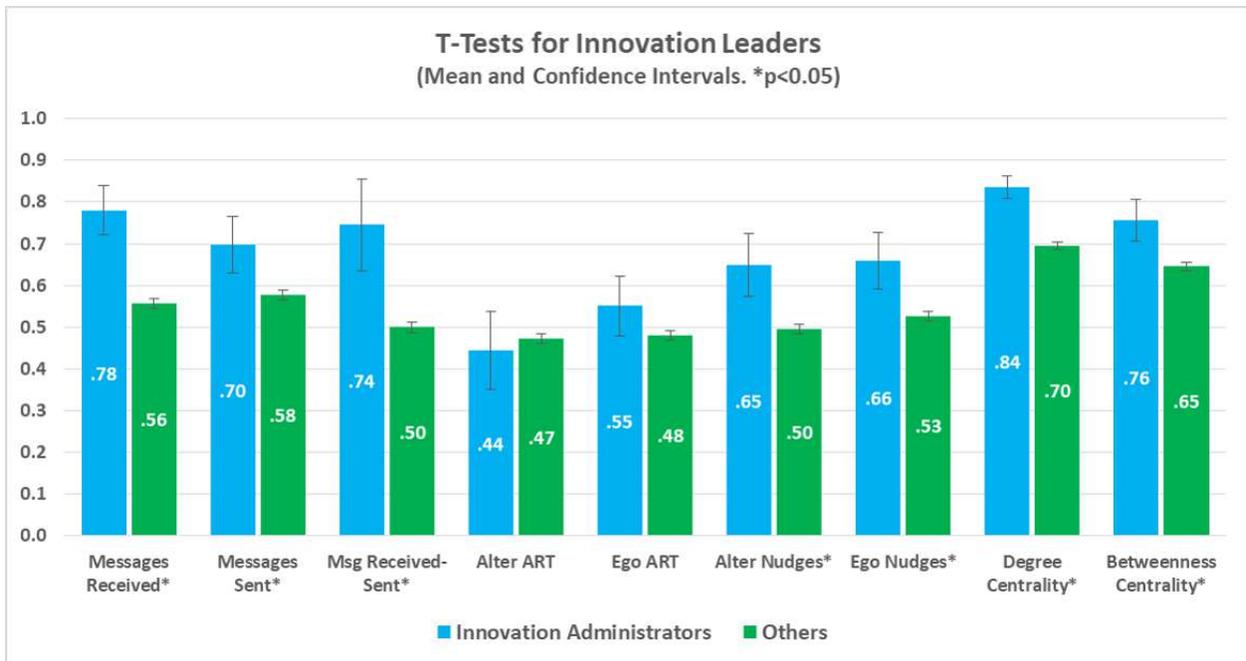

**Figure 2.** Significant T-tests (* p < .05).

Based on previous studies that differentiated intrapreneurship behaviors of individuals based on their role (Damanpour, 1988; Dyer et al., 2011; Gawke et al., 2019), we highlighted the role of innovation administrators, since their job role is not directly comparable to award-oriented and product-oriented innovators. Innovation administrators often act as mentors for their colleagues sharing knowledge and advice and coordinate and supervise bigger innovation projects; award- and product-oriented innovators, on the other hand, have a more operational role, being directly involved in research activities, as part of the teams supervised by the innovation administrators.

Our next models indicate a difference in communication behaviors of award-oriented innovators from product-oriented innovators. We used several ANOVA models to find the significant differences (group mean) in the communication behavior between product-oriented innovators, award-oriented innovators and their colleagues. As illustrated in Figure 3, significant ANOVA results ($p < .05$) are marked with an asterisk next to the variables' names. To identify the significant differences among groups we looked at the Tukey's HSD tests as illustrated in Table 3.

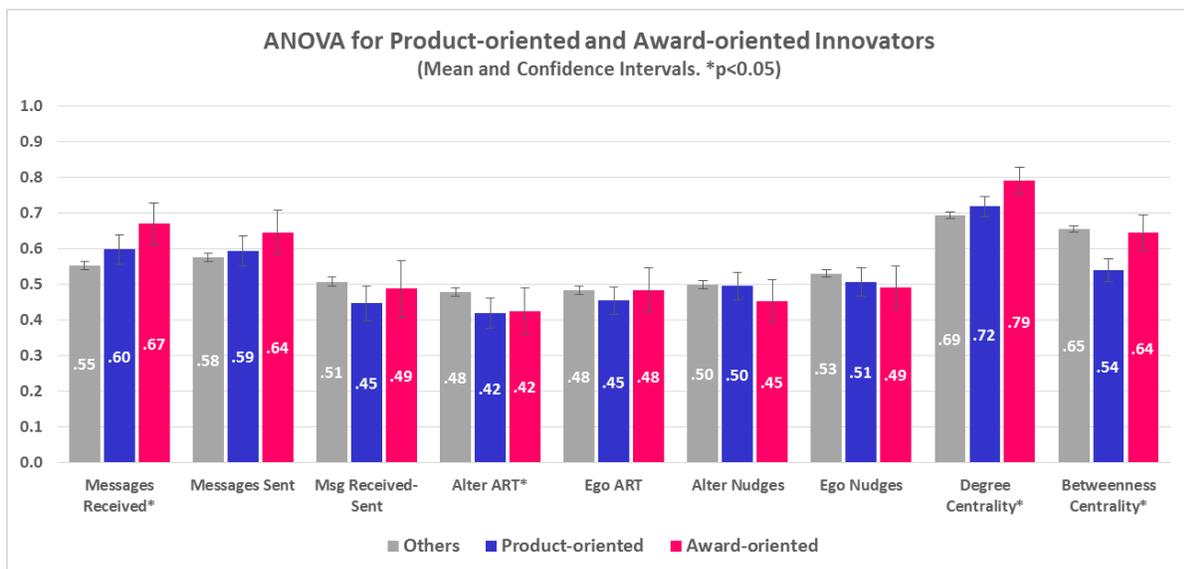

**Figure 3.** ANOVA for product-oriented and award-oriented innovators vs others (* p < .05).

Degree centrality and betweenness centrality are significantly higher for award-oriented innovators, both when compared with the product-oriented innovators and with other researchers. It seems that innovators who won the business award were more deeply embedded in the email network, also showing more bridging ties, within and outside the organizational boundaries. They also received more messages, with respect to other researchers. This is aligned with previous studies showing that innovators are positioned in the core of their social network, with similar values of messages sent and received (Gloor et al., 2008; Wen, Gloor, Fronzetti Colladon, Tickoo, & Joshi, 2019). The communication pattern of product-oriented innovators is characterized by lower values of Alter ART. This indicates that on average others take less time to respond to the e-mails of product-oriented innovators. This could be attributable to the higher respect they get from others, which makes others answer faster to their e-mails. It could also be founded in the collaborative nature of their work: since they are very active in publishing and applying for patents, they might need to rely on teamwork and are dependent on receiving timely contributions from others to complete their task.

| Variable | | Sum of Squares | df | Mean Square | F | Sig. | Mean | Post hoc analysis (Tukey HSD) | | |
|---|---|---|---|---|---|---|---|---|---|---|
| | | | | | | | | Other Researchers | Product Innovator | Award Innovators |
| Messages Received | Between groups | .911 | 2 | .455 | 7.520 | .001 | Others = .55 | | | ** |
| | Within groups | 117.507 | 1941 | .061 | | | ProductInn = .60 | | | |
| | Total | 118.418 | 1943 | | | | AwardInn = .67 | ** | | |
| Messages Sent | Between groups | .281 | 2 | .140 | 2.375 | .093 | Others = .58 | | | |
| | Within groups | 114.773 | 1941 | .059 | | | ProductInn = .59 | | | |
| | Total | 115.054 | 1943 | | | | AwardInn = .64 | | | |
| Msg Received - Sent | Between groups | .458 | 2 | .229 | 2.702 | .067 | Others = .51 | | | |
| | Within groups | 164.618 | 1941 | .085 | | | ProductInn = .45 | | | |
| | Total | 165.077 | 1943 | | | | AwardInn = .49 | | | |
| Alter ART | Between groups | .559 | 2 | .280 | 4.506 | .011 | Others = .48 | | * | |
| | Within groups | 120.489 | 1941 | .062 | | | ProductInn = .42 | * | | |
| | Total | 121.048 | 1943 | | | | AwardInn = .42 | | | |
| Ego ART | Between groups | .109 | 2 | .054 | .894 | .409 | Others = .48 | | | |
| | Within groups | 117.791 | 1941 | .061 | | | ProductInn = .45 | | | |
| | Total | 117.899 | 1943 | | | | AwardInn = .48 | | | |
| Alter Nudges | Between groups | .107 | 2 | .053 | .961 | .383 | Others = .50 | | | |
| | Within groups | 107.595 | 1941 | .055 | | | ProductInn = .50 | | | |
| | Total | 107.702 | 1943 | | | | AwardInn = .45 | | | |
| Ego Nudges | Between groups | .138 | 2 | .069 | 1.206 | .300 | Others = .53 | | | |
| | Within groups | 110.695 | 1941 | .057 | | | ProductInn = .51 | | | |
| | Total | 110.833 | 1943 | | | | AwardInn = .49 | | | |
| Degree Centrality | Between groups | .557 | 2 | .278 | 7.894 | .000 | Others = .69 | | | ** |
| | Within groups | 68.471 | 1941 | .035 | | | ProductInn = .72 | | | * |
| | Total | 69.028 | 1943 | | | | AwardInn = .79 | ** | * | |
| Betweenness Centrality | Between groups | 1.624 | 2 | .812 | 18.099 | .000 | Others = .65 | | *** | |
| | Within groups | 87.064 | 1941 | .045 | | | ProductInn = .54 | *** | | ** |
| | Total | 88.688 | 1943 | | | | AwardInn = .64 | | ** | |

*Note.* *p < .1; **p < .05; ***p < .01. ProductInn = Product Innovators; AwardInn = Award Innovators; Others = Other Peers.

**Table 3.** ANOVA and Tukey HSD post hoc tests.

Figure 4 summarizes our results and shows which indicators were positively or negatively associated to different types of innovators.

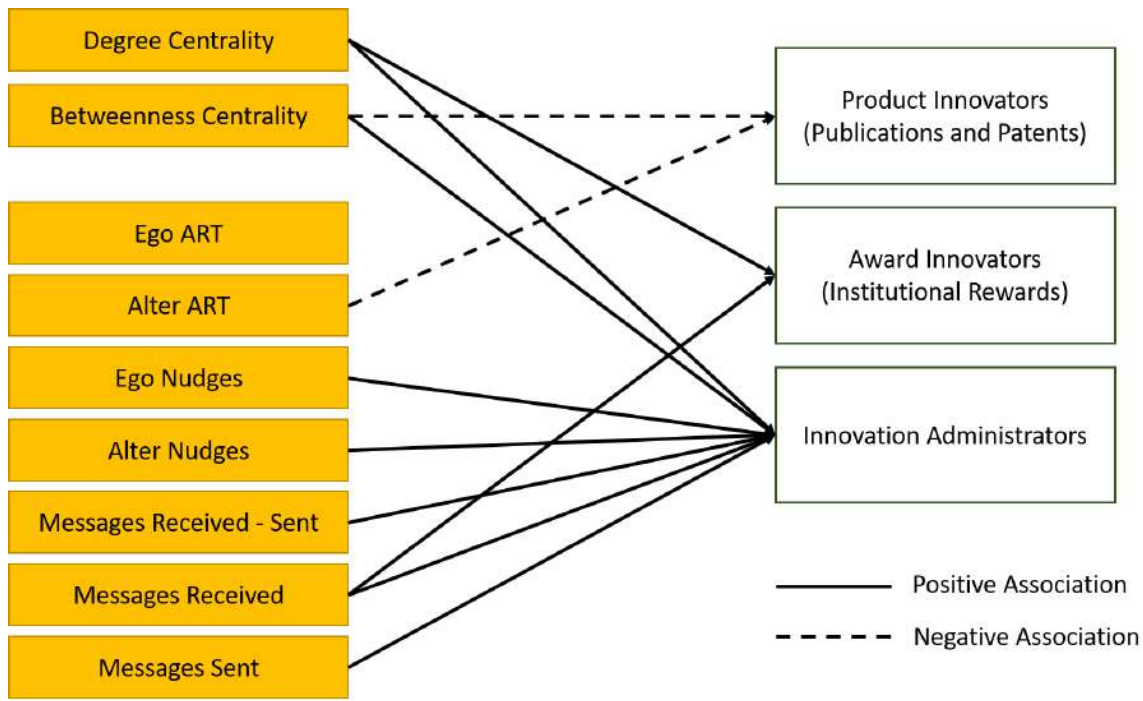

**Figure 4.** Summary of Results.

Researchers holding an administrative role had a distinctive online communication style: they were more central among their peers interacting with a higher number of peers (H1), and they sent more emails/reminders to others to solicit a response (higher ego nudges), showing a higher level of engagement in conversations (H4). In addition, they received more messages than they sent, probably because of their expertise and the need for others to access their knowledge (H5). Their role is to offer suggestions, ideas and support to award-oriented and product-oriented innovators. They are the point of reference for other researchers, based on their background and position in the institutional hierarchy (they usually have higher ranks). Their role as information hubs was only partially supported (H2). This could be explained by their role as subject matter experts, which requires them to respond directly to queries, and coordinate the process from an administrative and operational standpoint. A potential explanation for the high number of people with whom innovation administrators exchanged emails is that their responses to questions were often included in multiple communications with single messages containing many content-related responses. By contrast, innovation administrators did not show significant differences with respect to average response times; therefore, H3 is not supported for this specific sub-sample. In general, the Average Response Time was not significantly associated with an innovative behavior also for the other two types of innovators. This is probably due the fact that the volume of incoming email affects behaviors of recipients and the length of time it takes to reply (Soucek & Moser, 2010). Researchers who are focused on publishing, designing their products or submitting their proposals need time to study, reflect and connect with other people (not necessarily online), so their attention might be spread across many tasks, increasing the time they respond to emails. This is consistent with some studies that have discussed the role of time and time management as related to creativity. Creative individuals who are able to manage their mental time have a higher chance to meet their creative goals (Zampetakis, Bouranta, & Moustakis, 2010). Creativity and innovation require time, as it takes time to connect one idea to the other, and to find remote associations. Zampetakis et al. (2010) showed that individual creativity is significantly related to time management behaviors (daily planning, and confidence on long-range planning) and time attitudes (perceived control of time, tenacity and preference for disorganization). Other studies highlighted the importance of allocating adequate time for creative processes, specifically to understand and reframe the problem (Ronen, Goldschmidt, & Erez, 2018) and illustrate how high levels of time pressure can hamper creative processes, while moderate levels can stimulate team member creativity through motivational and cognitive processes (Khedhaouria, Montani, & Thurik, 2017).While we expected to find no significant differences between the communication patterns of product-oriented innovators and award-oriented innovators (H6), the analysis showed that we might be able to recognize different types of innovators from the way they interact online. Specifically, award innovators, mostly concerned with internal recognition, were much more central in the email networks: they received more messages, exchanged emails with a higher number of contacts, and acted more as information brokers. These findings are aligned with previous literature that shows how innovators tend to be more central (Becker, 1970) and are good at networking with a diversity of stakeholders (Dyer et al., 2011). A possible explanation of why award innovators create more ties and interact with a higher number of people via email is the need to build political capital, increase awareness and improve the chances for their ideas to be recognized and awarded. The more they spread the word about their ideas, the more awareness they create and the higher the chance to be "talked about" in meetings and among colleagues. As Kanter (1988) mentioned with reference to the administrative and technical innovation processes, "*corporate entrepreneurs often have to pull in what they need for their innovation from other departments or areas, from peers over whom they have no authority and who have the choice about whether or not to ante up their knowledge, support, or resources, to invest in and help the innovator*". The same explanation could be used to describe the results in terms of messages received, where there is a significant difference between award-oriented innovators and other peers. Innovators recognized with awards tend to receive more emails than other researchers, while product-oriented innovators do not significantly differ from the other two groups. Obtaining organizational recognition requires specific behaviors and actions that might differ from an innovative output represented by publishing a scholarly article or submitting a patent application. Award oriented innovators need to be more active showcasing their invention and emphasizing details that are easily understood by a broad audience. Both types of innovators seem to have central roles in their respective networks and tend to be well respected as one might deduce from the lower alter ART – as people are quite fast at responding to their emails.

While the success of an entrepreneurial activity involves similar processes such as building and leveraging social capital independently of the type of innovation involved (Ibarra et al., 1993), the specific information and resources that innovators are trying to identify in their networks might vary. This would explain why award-innovators have higher degree and betweenness centrality and command more control on the flows of information going through them: in order for their idea to be awarded, they need to connect with stakeholders outside of their unit or organization, to build recognition, awareness and external support.

## 5. Conclusions

This study extends innovation and creativity research both at the individual level of analysis and in the domain of informal networks (Gawke et al., 2019; Keller, 2017; McKay & Kaufman, 2019). We investigated the association of organizational recognition (institutional awards) and external/peer recognition (patents and publications), with network centrality and interactivity in the communication patterns of researchers. Our study offers an important contribution to the understanding of the types of communication behavior that characterize innovators motivated by different goals or tasked with different activities. To our knowledge, this is the first study where award-innovators and product-innovators are employed as distinguishing categories to classify separate innovation outcomes. On the theoretical side we introduce a novel method of identifying innovators by analyzing their email communication behavior. On the practical side, we discuss how accessing the corporate email archive will create invaluable insights for managers, allowing them to better understand, coach, and reward their most valuable asset, the capability to innovate of their employees.

## 5.1. Theoretical Implications

In line with the stream of literature discussing psychological traits and behaviors of creative individuals within organizations and in everyday life (Brem et al., 2016; Fürst & Grin, 2018; McKay & Kaufman, 2019), as well as with literature discussing the role of time management and interpersonal ties (Marrone, 2010), this study provides important new information on the communication patterns that are associated with creative individuals and innovators.

This study makes an important extension to innovation theories by focusing on indicators not previously examined in the literature. While innovation-relevant skills and traits have been extensively discussed (Amabile, 1988; Scott & Bruce, 1994; West & Richter, 2008), less attention has been focused on examining which communication patterns are indicative of innovative outcomes.

This paper contributes to the further understanding of factors impacting the creative process and specifically how the communication behavior can be analyzed to identify differences between innovators who are prolific regarding scientific publications or patents, and innovators who are driven by political and institutional recognition. In particular, being very central in their online communication networks might not be associated to a successful innovative outcome, as shown by product-oriented innovators who are in a non-central position. This is aligned with other studies that demonstrate how having too many weak ties or being too central may actually constrain creativity and innovation (Perry-Smith, 2006; Perry-Smith & Shalley, 2003).

Our finding on response time is aligned with recent empirical evidence (Tinguely et al., 2019) showing a positive association between the ability to stay focused when time pressure rises over time and employees' higher position within the organization as well as higher quality relationships with mentors. While Tinguely and colleagues (2019) used surveys to assess the role of time pressure on creativity, we chose an approach that analyzes online communications, reducing data collection biases. Our study provides a methodological confirmation of the benefits of research methods based on collecting e-mail archives (Gloor, Fronzetti Colladon, Giacomelli, Saran, & Grippa, 2017; Sivarajah, Kamal, Irani, & Weerakkody, 2017). Even though mining email archives can provide an "almost complete" view of R&D members' social network (Grippa, 2009), the virtual nature of today's innovative teams – with ties outside of organizational boundaries – makes this collection method reliable and promising. The methodology presented in this paper offers a complementary approach to assess creativity outcomes, which cannot be compressed into a single scale of measurement. As suggested by Sternberg (2018), measuring creativity requires the adoption of truly multidimensional and interdisciplinary measures. Allen and colleagues (2016) studied the communication of startup research labs and proved the advantages of being embedded in the communication network core, with little or no impact of geographical locations: interactions induced by physical proximity of employees do not necessarily lead to knowledge exchange and innovation. Emails create a virtual workspace which has no time constraints and can cross the physical boundaries of business departments, allowing collaboration on a global scale. For example, a longitudinal study on knowledge creation by members of a virtual creative design team found that the team was extremely successful and innovative, in spite of the lack of frequent informal face-to-face interaction (Majchrzak, Rice, King, Malhotra, & Ba, 2000).

### 5.2 Managerial Implications

This study has important practical implications as our results can help managers to identify hidden innovators and people who could potentially play this role, by looking at their communication styles. As suggested in the Four C model of Kaufman and Beghetto (2009) nearly all aspects of creativity can be experienced by nearly everyone. The method we used in this study could be replicated to promote self-awareness and reflective thinking among employees, especially within organizations that support corporate entrepreneurship. Managers can use a method similar to the one described in our study to identify employees with specific intrapreneurship traits and support them via resources that can bolster creativity and innovation (Gawke et al., 2019; S. Park, Lee, & Song, 2017). Human resource managers and senior organizational leaders are encouraged to design and develop feedback sessions to discuss how communication patterns can improve the dyadic relationships between employees and supervisors (Gloor, Fronzetti Colladon, Giacomelli, et al., 2017). This may give employees the levels of autonomy and discretion necessary for innovation to emerge (Dyer et al., 2011; Oldham & Cummings, 1996; Scott & Bruce, 1994). Support for autonomy, defined as managers' understanding and acknowledgement of their subordinates' perspectives, encouraging self-initiation and minimizing pressures and controls, is usually associated with improved outcomes (Gagné & Deci, 2005; Mirchandani & Lederer, 2008). Work environments that support autonomy facilitate internalization of extrinsic motivation, resulting in increased motivation.

As suggested by previous literature, part of the innovator's DNA is to identify and test ideas through a network of diverse individuals who can extend their perspectives and increase their knowledge (Dyer et al., 2011). Our study confirms previous empirical evidence showing how innovators develop more contacts with others who can support the creativity process (Amabile, 1988; Dyer et al., 2011; Perry-Smith, 2006; Ray et al., 1997). This has implications for managers as more initiatives devoted to establish and nurture collaborative networks of researchers should be implemented.

Our work additionally contributes to the discussion on career advancement, and in particular to understanding which competencies and behaviors are differentiating candidates for promotion (Chong, 2013): what happens to communication patterns in the social network once innovators advance in their careers from the more basic role of product-oriented innovators, to award-winning and leadership positions? Our study offers empirical evidence about the online communication styles that differentiate innovators from other R&D employees and administrators. It has practical implications in guiding R&D managers in recognizing and nurturing innovators within their department. The identification of current and untapped expertise is invaluable when setting up research teams and managing talent (Wen et al., 2019). Understanding the way innovators communicate can help managers support the kind of communication and interaction dynamics that foster innovation. Figure 5 shows some tips for managers, in terms of social behavior and time management, derived from our findings.

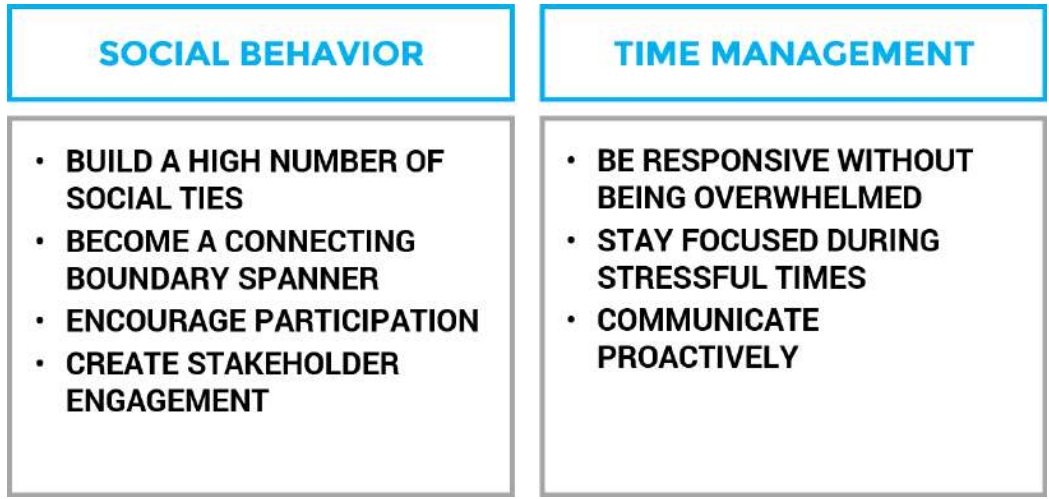

**Figure 5**. Managerial Implications.

### 5.3 Limitations and Future Research

Because of data limitations, we were not able to consider possible mediating mechanisms such as personality traits and individuals' involvement in the larger scientific and industrial community. The nature and length of the relationships with industrial partners and other researchers is an important mediator that could influence the structural properties of the communication network. Similarly, individual factors such as extraversion or openness to experience are important mediators that should be included in future studies to fully measure creativity and innovation (Amabile, 1988; Fürst & Grin, 2018). Given the importance of individual factors and personality traits impacting creativity and innovation (Amabile, 1988; Brem et al., 2016; Ray et al., 1997), we strongly encourage future research that combines the analysis of these traits with that of online communication.

Another limitation of this study is the lack of control for the content of the email exchange (due to privacy agreements), and the lack of other control variables such as the age and gender of employees, their past performance, and the length of their tenure in the company. Variables such as tenure and past performance could influence actors' social positions (Wen et al., 2019), i.e. their network centrality. Past performance, however, is at least partially reflected in employees' rank, which is the control variable we use in our models. The lack of information about the tenure in the company is mediated through the culture of the company, which promotes a work environment for life, i.e. most employees have been with the company for decades. In addition, it would be interesting to replicate our experiment in other business contexts, and in other countries, to see whether our findings depend on culture or on business activities (Oldham & Cummings, 1996; Rinne et al., 2013). The approach we presented could be tested in other companies and potentially recalibrated according to the characteristics of different business settings. The focus of this research is on the identification of communication patterns that distinguish different kinds of innovators from their peers. We do not claim to prove causality, for which we advocate dedicated future research.

Another area of investigation would be the inclusion of email content in the analysis, which could offer deeper insights into the profiling of innovators. By leveraging the progress in big data analytics, we encourage the adoption of methodologies of semantic analysis and sentiment analysis to calculate other language-related measures and understand how the emotion we transfer in our online communication, or the complexity of our language, impact the creative output (Gloor, Fronzetti Colladon, Giacomelli, et al., 2017).